\documentclass[aps,physrev,preprint,groupedaddress]{revtex4-2}
\usepackage{amsmath}
\usepackage{graphicx,epstopdf}
\usepackage{color, colortbl}
\usepackage{caption}
\usepackage[normalem]{ulem}
\usepackage{subcaption}
\newcommand{\eq}[1]{{Eq. (\ref{#1})}}
\DeclareMathOperator{\e}{e}

\begin{document}


\title{Impurity localization, and collision properties of symbiotic dark-bright solitons in  superfluid-impurity  system}


\author{Dileep K}
\email[]{dileepk.17@res.iist.ac.in}
\affiliation{Department of Physics, Indian Institute of Space Science and
Technology, Thiruvananthapuram, 695 547, India}

\author{S Murugesh}
\email[]{murugesh@iist.ac.in}
\affiliation{Department of Physics, Indian Institute of Space Science and
Technology, Thiruvananthapuram, 695 547, India}


\begin{abstract}
We investigate the dynamics of a binary mixture of Bose-Einstein condensates in the impurity limit-- where one component is dilute enough to be treated like an impurity-- and confined to two dimensions. Using the mean-field coupled Gross-Pitaevskii equations, we find that the binary mixture supports the formation of stable symbiotic dark-bright solitons when the inter- and intra-component interactions are repulsive. We further study the interaction between solitons and observe that the solitons undergo merging and repulsion depending on the relative phase between the bright component of the composite structure.
\end{abstract}


\maketitle

\section{Introduction}
Localized excitations in nonlinear dispersive media 
are commonly encountered in a variety of physical systems --- propagation of high intensity light waves in dielectric media, waves in fluids and superfluids, Bose-Einstein condensates (BECs), spin waves in magnetic materials, energy propagation in certain biomolecules such as DNA, to name a few~\cite{kevrekedisbook,kosevich, scott, dauxois}. It is fairly well understood that such mobile and stable localized excitations are witnessed in non-linear media on account of a delicate balance between the dispersive nature of the medium of propagation and self steepening due to its nonlinear response. A vast body of exact results exists explaining such behavior in  one-dimensional systems, governed by fundamental models such as Korteweg-De Vries equation (KDV), the modified KDV, Sine-Gordon equation, Nonlinear Schr\"{o}dinger equation (NLSE), Toda lattice model, etc,  and their extended higher order versions~\cite{ablowitz1981}. Whereas several of these aforementioned models are {\it integrable}, and hence supplemented by an infinite number of constraints, stable localized excitations are nevertheless witnessed in near integrable and driven non-linear systems as well, wherein they exhibit considerable robustness under collisions and perturbations. Along with the balance between dispersion and self steepening, driven systems can also find an additional balance with dissipation inherent in the media, 
often leading to solitonic excitations~\cite{akhmediev2005}. In higher dimensions however, the stability of such localized excitations are not directly carried over. Dispersion in higher dimension often proves stronger to collapse the balance --- dimension usually plays a detrimental role to stability of localized excitations in nonlinear propagation. Instead, nonlinear systems in higher dimensions are host to other non-trivial excitations with an inherent  geometric structure, such as vortices, for instance~\cite{madison, fetter}.

One way to sustain localizations in higher dimensions is to generally go beyond the standard self-focusing NLSE by including higher-order effects, which effectively balance the transverse dispersion and stabilize two-dimensional solitons. For instance, in nonlinear optics, replacing Kerr nonlinearity by saturable nonlinearity can suppress the transverse instabilities and generate 2D solitons that are stable for longer duration~\cite{yang, kivshar2003}. However, these solutions are not indefinitely stable, since the transverse instabilities are only suppressed and not completely removed.

It should be noted that, {\it propagation} is an additional balancing factor, besides dispersion and self steepening, in bringing about stable localized excitations such as solitons. In higher dimension, however, while the dispersion is thus compensated by self-steepening {\it only along the direction of propagation}, the dispersion along the transverse direction cannot be balanced by self-steepening alone. In this paper we consider a binary mixture of BECs, wherein the interaction between the two components provides the requisite additional balance, apart from nonlinearity, to create stable localized and traveling excitations in both components. Since the interaction between the two components is critical for stability, the localization is {\it symbiotic} - i.e., they occur together at the same point in space and time, one dark and the other bright. 

Atomic BECs provide a highly tunable experimental platform for investigations of nonlinear phenomena. In the mean-field limit, the dynamics of the condensate is described by the Gross-Pitaevskii equation (GPE), which is an NLSE. The coefficient of nonlinearity in
the GPE specified by the s-wave scattering length can be tuned in experiments by means of 
magnetically~\cite{inouye, cornish} and optically~\cite{fatemi, blatt} induced Feshbach resonances. Moreover, the trapping potentials, engineered using lasers and magnetic fields, can effectively reduce the dimensionality of the system, enabling a practical realization of one- and two-dimensional NLSEs. The effective 1-D GPE supports dark and bright solitons depending on the sign of interparticle interaction~\cite{burger,denschlag,khaykovich}. In 2D, the GPE predicts the existence of topological excitations such as vortices~\cite{dalfovo1996}, which have been experimentally observed in trapped atomic BECs~\cite{madison, matthews}.

BECs also offer a versatile setting for investigating multicomponent generalizations of the NLSE. Such mixtures have been experimentally realized~\cite{myatt} and theoretically studied for their ground state~\cite{ho,pu, esry} and low-energy excitations~\cite{busch, graham, pu98, esry98}. The intercomponent interactions in these mixtures influence the equilibrium and dynamic properties of the BECs. For instance, a binary mixture of BECs may undergo phase separation depending on the strength of interspecies interaction~\cite{lee, hall}. In one-dimension, the binary mixtures can also support symbiotic structures such as dark-dark~\cite{ohberg} and dark-bright~\cite{Busch2001} solitons. 

An interesting limit of the binary mixture is the case where one component is dilute such that the majority component constitutes a background condensate density with which the small fraction of atoms– referred to as {\it impurities}– interact via interspecies coupling~\cite{catani}. The impurities can be atoms of the same type in different hyperfine states or different atoms. If the concentration of impurity atoms is very low, then the impurities can move around in the Bosonic environment, leading to the formation of a quasiparticle called Bose polaron~\cite{jorgensen, hu2016}. The interspecies interaction in these systems has been shown to induce self-localization of neutral impurity atoms~\cite{kalas, sacha}. Such localized impurity states  in  binary BECs have been investigated using the 1D mean-field model for attractive interactions in ~\cite{Edmonds_2019}, wherein collision and dimer formation with respect to density variations  have been investigated in detail. In 2D, the confinement and manipulation of impurity atoms by superfluid vortices has been studied in~\cite{edmonds}.


In this paper we investigate a coupled system consisting of a mixture of 2-D confined Bose condensed atoms in states given by wave functions $\psi_1$ and $\psi_2$.
The wave functions represent condensed atoms in two distinct states, with a vast majority in state 
$\psi_1$, comparatively - i.e., $\int{|\psi_1|^2~d{S}} >> \int{|\psi_2|^2~d{S}}$. Alternately, $\psi_2$ could  equally be  considered to represent a small distribution of impurity atoms in a condensate that is predominantly in state $\psi_1$. Experimentally, such a binary model can be realized either by controlled doping of impurities~\cite{spethmann2012,spethmann, hohmann2015} or by transferring a small fraction of atoms to another hyperfine state~\cite{Aycock}. In our model, we have considered both the inter and intra species interactions to be repulsive. Interestingly, as we show below, the impurity species ($\psi_2$) shows a distinct localization, while the primary condensate component ($\psi_1$) shows a symbiotic near zero dip in its complex amplitude. The two excitations propagate together, akin to a dark-bright soliton pair. As we argue below, the pairing is critical for the stability of both excitations, which is otherwise not possible, given the dimensionality.  

Further, encounters between such pulses result in either repulsive collisions, or the pulses coalesce, determined by the relative phases between the colliding pulses. Such phase determined repulsive and attractive collisions are fairly well known in nonlinear media, particularly in optical pulse propagation~\cite{george}. Yet coalescence is, to our knowledge, fairly new. 
 A suspicion that the dark pulses of the primary condensate could be vortices, is set aside by their phase profile 
and a non-vanishing condensate wavefunction.
We also present a qualitative argument on how such localization is possible in this system where all interactions are primarily repulsive, and in spite of its higher dimensionality. 

In Section~\ref{sec:2}, we describe the model, 2-D coupled Gross-Pitaevskii equations, in detail. Section~\ref{sec:3} discusses the stability of the homogeneous state of the binary mixture using linear stability analysis. The complete nonlinear evolution of the perturbed homogeneous solutions is discussed in Section~\ref{sec:4}, where we observe impurity localizations accompanied by dips in the superfluid density. Section~\ref{sec:4} also discusses the controlled generation of symbiotic dark-bright solitons in the binary system and their collisions. In section~\ref{sec:5} we summarize the findings of our study.

\section{2-dimensional condensate with impurities }
\label{sec:2}
We consider a planar binary mixture of Bose-Einstein condensed atoms of the same species in two different internal states represented by the macroscopic wave functions $\psi_1$ and $\psi_2$. The mixture is assumed to be prepared in such a way that the condensate is predominantly in state $\psi_1$, while only a small fraction of superfluid atoms is transferred into state $\psi_2$. The controlled addition of impurities to the condensate has been experimentally demonstrated in $^{87}$Rb BEC by applying an rf pulse, which transfers a small fraction of atoms to another hyperfine state~\cite{Aycock}. Consequently, the condensate may be regarded as a system of superfluid atoms collisionally coupled to  impurity atoms of the same type. The dynamics of such a mixture is then described by the coupled 2-D Gross-Pitaevskii equations (GPEs)~\cite{akram, edmonds, Edmonds_2019, Aycock, bhat},
\begin{subequations}
\begin{align}
i \hbar\frac{\partial \psi_1}{\partial t} &= \left(-\frac{\hbar^2}{2m} \nabla^2+U_{11}|\psi_1|^2+U_{12}|\psi_2|^2 \right)\psi_1, \label{eq:1a}\\
i \hbar\frac{\partial \psi_2}{\partial t} &= \left(-\frac{\hbar^2}{2m}\nabla^2+U_{12}|\psi_1|^2 \right)\psi_2, \label{eq:1b}
\end{align}
\label{eq:1}
\end{subequations}
where $U_{11}$ denotes the effective interaction between atoms in state $\psi_1$, while $U_{12}$ is the interaction between superfluid and impurity atoms. Note that the interaction between the impurity atoms has been neglected in \eq{eq:1b} on account of its diluteness. The coefficients $U_{11}=\frac{\sqrt{8\pi}\hbar^2a_{11}}{ma_z}$ and $U_{12}=\frac{\sqrt{8\pi}\hbar^2a_{12}}{ma_z}$ are determined by the respective 3D scattering lengths $a_{11}$, $a_{12}$, and the harmonic oscillator length $a_z=\sqrt{\frac{\hbar}{m\omega_z}}$, where $\omega_z$ is the frequency of the harmonic trapping potential.

In the following, we investigate the dynamics of the coupled system described by~\eq{eq:1}, by converting them to a dimensionless form. Upon rescaling the variables $x\rightarrow x/a_z$, $y \rightarrow y/a_z$, $t \rightarrow \omega_z t$ and $\psi_j \rightarrow \psi_j a_z$, Eq.~(\ref{eq:1}) can be written in the dimensionless form

\begin{subequations}
\begin{align}
i \frac{\partial \psi_1}{\partial t} &= \left(-\frac{1}{2} \nabla^2+g_{11}|\psi_1|^2+g_{12}|\psi_2|^2 \right)\psi_1, \label{eq:2a}\\
i \frac{\partial \psi_2}{\partial t} &= \left(-\frac{1}{2}\nabla^2+g_{12}|\psi_1|^2 \right)\psi_2, \label{eq:2b}
\end{align}
\label{eq:2}
\end{subequations}
where $g_{11}=\frac{\sqrt{8\pi}a_{11}}{a_z}$ and $g_{12}=\frac{\sqrt{8\pi}a_{12}}{a_z} $ are the dimensionless coupling constants. The 1D version of \eq{eq:1} has been studied for impurity localization in BEC~\cite{akram}, modulation instability~\cite{bhat} and impurity transport in the attractive regime~\cite{Edmonds_2019}. In 2D, the equation has been shown to support impurity localizations that are trapped by vortex solutions in the superfluid component~\cite{edmonds}.

\section{Stationary states and their stability}
\label{sec:3}
We first discuss the dynamics of stationary states of the binary mixture, where the density of each component is homogeneous in space, $n_j=|\psi_j|^2$. Since the number of particles in each state is conserved, equilibrium properties of the condensate can be described by introducing the chemical potentials $\mu_1$ and $\mu_2$. The wavefunctions for the stationary states are thus of the form
\begin{equation}
\psi_j = \sqrt{n_j} \e^{-i\frac{\mu_j}{\hbar}t}.
\end{equation}
Substituting in \eq{eq:1}, we get $\mu_1 = U_{11}n_1+U_{12}n_2$, and $\mu_2= U_{12}n_1$. 

The homogeneous state in the binary mixture can be unstable when subjected to perturbations. These instabilities can often lead to localizations in the condensate, similar to soliton excitations. In this section, we review the stability of the condensate mixture using linear stability analysis and determine the conditions under which the mixture is modulationally unstable~\cite{bhat}. We consider perturbations to the homogeneous state in the form
\begin{equation}
\psi_j = (\sqrt{n_j}+\delta \psi_j) \e^{-i\mu_j t},
\end{equation}
where the complex field $\delta \psi_j$, is the perturbation to the component $j$. Substituting in~\eq{eq:2}, we get to first-order in $\delta \psi_j$,

\begin{subequations}
\begin{align}
i \frac{\partial \delta \psi_1}{\partial t} &= -\frac{1}{2} \nabla^2 \delta \psi_1 +g_{11} n_1 (\delta \psi_1+\delta\overline{\psi_1})+g_{12}\sqrt{n_1n_2}(\delta \psi_2+\delta\overline{\psi_2}), \label{eq:3a}\\
i \frac{\partial \delta \psi_2}{\partial t} &= -\frac{1}{2}\nabla^2\delta \psi_2+g_{12}\sqrt{n_1n_2}(\delta \psi_1+\delta\overline{\psi_1}). \label{eq:3b}
\end{align}
\label{eq:3}
\end{subequations}
As in the 1-D case, the coupled linear equations can be solved by assuming plane wave solutions of the form $\delta \psi_j = a_j \e^{i(\mathbf{k}\cdot\mathbf{r}-\Omega t)} + b_j \e^{-i(\mathbf{k}\cdot\mathbf{r}-\Omega t)}$, where $\mathbf{k} = k_x \hat{x} + k_y \hat{y}$, is a real wave vector with magnitude $k=\sqrt{k_x^2+k_y^2}$. Equation~\ref{eq:3}, therefore reduces to the following set of linear homogeneous equations in the variables $(a_1, \overline{b}_1, a_2, \overline{b}_2)$
\begin{subequations}
\begin{align}
\left(\Omega-\frac{1}{2}k^2-g_{11}n_1\right)a_1-g_{11}n_1\overline{b}_1-g_{12}\sqrt{n_1n_2}(a_2+\overline{b}_2) &= 0,\\
-\left(\Omega+\frac{1}{2}k^2-g_{11}n_1\right)\overline{b}_1-g_{11}n_1a_1-g_{12}\sqrt{n_1n_2}(a_2+\overline{b}_2) &= 0,\\
\left(\Omega-\frac{1}{2}k^2\right)a_2 - g_{12}\sqrt{n_1n_2}(a_1+\overline{b}_1) &= 0,\\
-\left(\Omega+\frac{1}{2}k^2\right)\overline{b}_2 - g_{12}\sqrt{n_1n_2}(a_1+\overline{b}_1) &=0,
\end{align}
\end{subequations}
which has a nontrivial solution only when the dispersion relation
\begin{equation}
\Omega^2_{\pm} = \frac{k^2}{2}\left(\frac{k^2}{2}+g_{11}n_1\left(1\pm\sqrt{1+\frac{4g^2_{12}n_2}{g^2_{11}n_1}}\right)\right),
\label{eq:dispersion}
\end{equation} 
is satisfied. It is evident that the homogeneous state is stable only when the frequencies of the continuous modes are real, i.e., $\Omega^2_{\pm}>0$. Conversely, if one of the eigenfrequencies, $\Omega^2_{\pm}<0$, the perturbations will grow at the expense of the background homogeneous field. In fact, it can be directly verified from~\eq{eq:dispersion} that the homogeneous state is unstable regardless of the nature of interactions: When the intercomponent interaction between superfluid atoms is attractive ($g_{11}<0$), $\Omega_+$ perturbations will grow if the wavenumbers are in the range $0<k<\sqrt{2|g_{11}|n_1\left(1+\sqrt{1+\frac{4g^2_{12}n_2}{g^2_{11}n_1}}\right)}$, causing the homogeneous mode to be unstable, while for $g_{11}>0$, $\Omega_-$ frequencies are unstable for wavenumbers in the range $0<k<\sqrt{2g_{11}n_1\left(-1+\sqrt{1+\frac{4g^2_{12}n_2}{g^2_{11}n_1}}\right)}$. Remarkably, the homogeneous state is unstable with respect to periodic modulations when both inter- and intra-component interactions are repulsive, i.e., when $g_{11},g_{12}>0$. This is in contrast to the scalar BECs with repulsive interactions, which are modulationally stable~\cite{theocharis}.

\section{Impurity localisation and symbiotic soliton-like excitations}
\label{sec:4}
As mentioned in the preceding section, the MI of the homogeneous state can lead to the formation of stable 2-D soliton structures which maintain their shape and velocity during propagation. In this section, we discuss the dynamics of soliton excitations in the superfluid-impurity system by numerically solving the coupled GPEs, Eq.~\ref{eq:2}. The simulations are performed using the split-step fourier method on a spatial domain of size $L \times L$ with lattice spacing $\Delta x = \Delta y = \frac{L}{N}$, where $L=20$ and $N=256$, and a time step $\Delta t =0.001$ (simulations performed using $N=512$ and $\Delta t =0.0001$ produced qualitatively identical results). We assume that the initial condition is a uniform mixture of superfluid and impurity atoms, seeded with small perturbations. In our simulations, the fraction of impurity atoms, $N_2/N$, in the binary mixture is varied between 0 and 0.06. Depending on the nature of interactions, the continuous wave solution evolves into 2-D localizations as described below.
\subsection{Repulsive interactions: Symbiotic solitons}
\begin{figure}
\centering
\begin{subfigure}{0.42\textwidth}
\centering
\resizebox{1\textwidth}{!}{\input{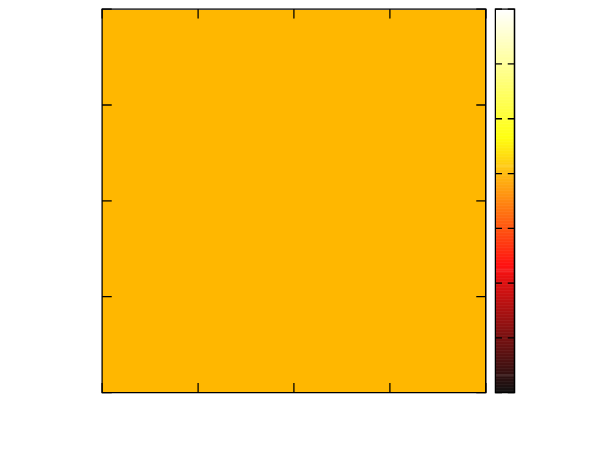}}
\caption*{(a)}
\label{fig:1a}
\end{subfigure}
\hspace{1cm}
\begin{subfigure}{0.42\textwidth}
\centering
\resizebox{1\textwidth}{!}{\input{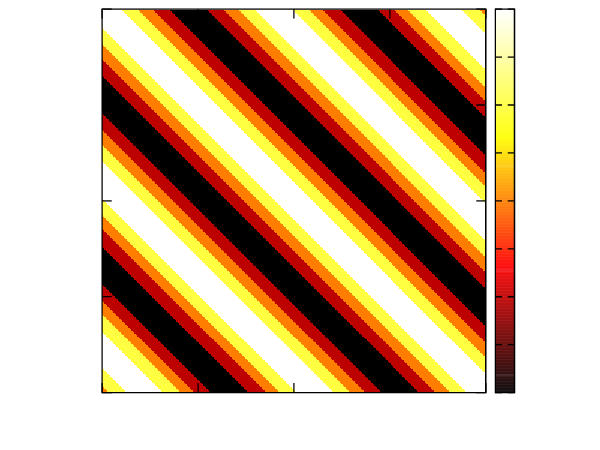}}
\caption*{(d)}
\label{fig:1d}
\end{subfigure}
\begin{subfigure}{0.42\textwidth}
\centering
\resizebox{1\textwidth}{!}{\input{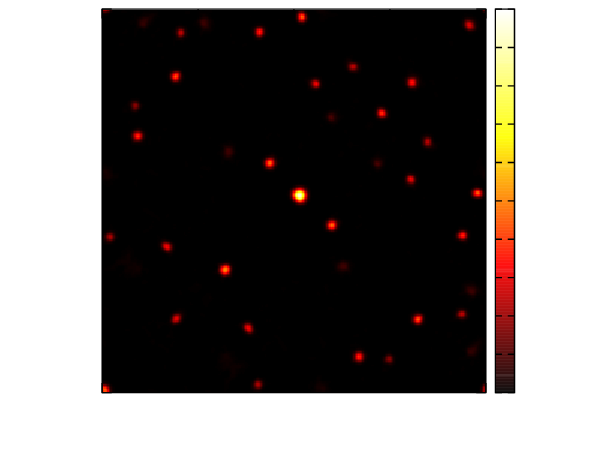}}
\caption*{(b)}
\label{fig:1b}
\end{subfigure}
\hspace{1cm}
\begin{subfigure}{0.42\textwidth}
\centering
\resizebox{1\textwidth}{!}{\input{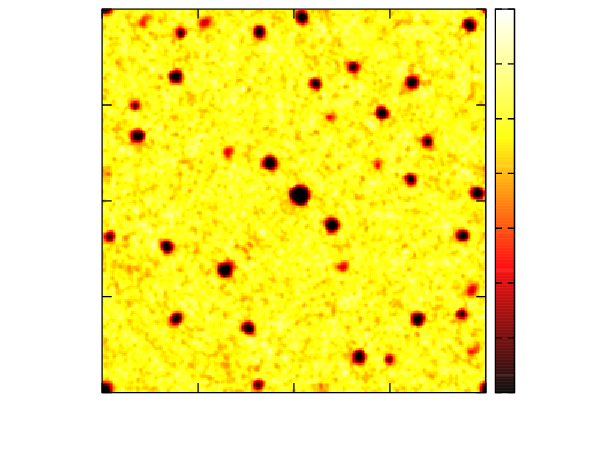}}
\caption*{(e)}
\label{fig:1e}
\end{subfigure}
\begin{subfigure}{0.42\textwidth}
\resizebox{1\textwidth}{!}{\input{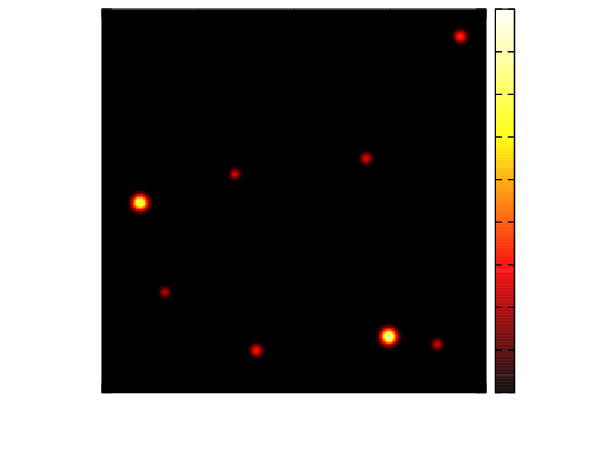}}
\caption*{(c)}
\label{fig:1c}
\end{subfigure}
\hspace{1cm}
\begin{subfigure}{0.42\textwidth}
\centering
\resizebox{1\textwidth}{!}{\input{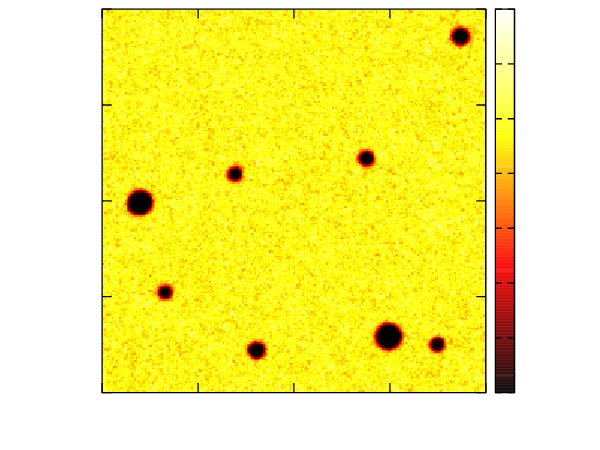}}
\caption*{(f)}
\label{fig:1f}
\end{subfigure}
\caption{ Dark-bright soliton formation in the superfluid-impurity system with $\frac{N1}{N}=0.94$ and $\frac{N_2}{N}=0.06$. The interaction strength $g_{11}=50$, and the ratio of interaction coefficients $\alpha=\frac{g_{12}}{g_{11}}=0.95$. (a)-(c) Time evolution showing the formation of bright solitons in the impurity component from the (a) initial homogeneous state. (d)-(f) Density depressions are observed in the corresponding intensity plot of the superfluid component.  The initial conditions are uniform density, seeded with small perturbations, for both components $\psi_1$ and $\psi_2$. See supplementary material for detailed animation of the time evolution.}
\label{fig:1}
\end{figure}

We consider the case in which the inter- and intra-component interactions are positive, i.e., $g_{11},g_{12}>0$. Repulsive contact interactions are known to stabilize dark solitons for scalar BECs. However, for the mixture of superfluid-impurity system, we observe that repulsive interactions result in the formation of impurity localizations in the form of bright solitons. Figure~(\ref{fig:1}(a)-(c)) shows the density maxima formed in the impurity component during the time evolution, for an impurity fraction $\frac{N_2}{N}=0.06$. The clustering of impurity atoms is accompanied by density depressions in the superfluid component, leading to phase separation of the binary mixture~(figure(\ref{fig:1})(d)-(f)). As we see below in Subsection C, when localized impurity pulses collide with one another, they either coalesce or repel depending on the relative phase between the colliding lumps --- coalesce when they have the same phase, and repel when the phases differ by $\pi$. Besides, the phase of the solitons also changes gradually as they propogate. This results in the impurity species witnessing a complete localisation eventually. In other words, with repulsive interactions, an initial miscible BEC undergoes phase separation characterized by the formation of {\it dark} and {\it bright} solitons. The number of generated localizations increases with the ratio of interaction strengths, $\alpha=\frac{g_{12}}{g_{11}}$. For a fixed value of $\alpha$, we find that the width of the soliton is determined by the superfluid interaction strength, $g_{11}$. As $g_{11}$ increases, the repulsion between superfluid atoms increases, leading to more localized soliton profiles.

It is important to note that localization of the impurity atoms is stabilized solely through repulsive contact interactions. Because the interaction between the impurity atoms is zero, the positive coupling coefficients ($g_{11},g_{12}>0$) induce an effective attractive interaction in the impurity component, which favors the formation of bright solitons.  We emphasize that the bright solitons observed here are not isolated solutions; rather, they are embedded in the density depressions of the superfluid component. Symbiotic coexistence of this type is commonly referred to as a {\it dark-bright} (DB) or more generally a {\it gray-bright} soliton~\cite{Busch2001,kevrekidis2004,kasamatsu2006}.

\subsection{Dynamics: A Single DB Soliton}
\begin{figure}[tb]
\centering
\begin{subfigure}{0.32\textwidth}
{\resizebox{1\textwidth}{!}{\input{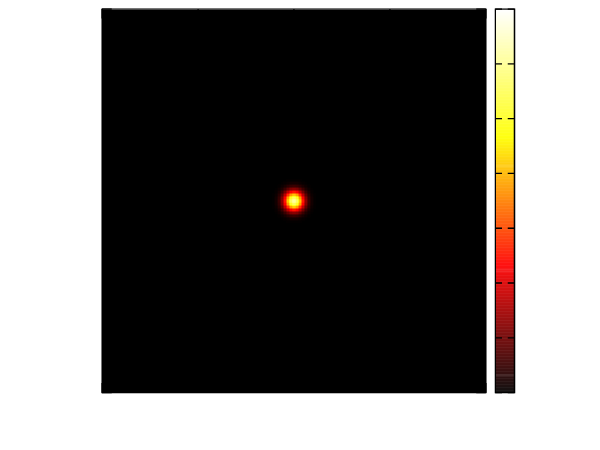}}}
\caption{}
\label{fig:2a}
\end{subfigure}
\begin{subfigure}{0.32\textwidth}
{\resizebox{1\textwidth}{!}{\input{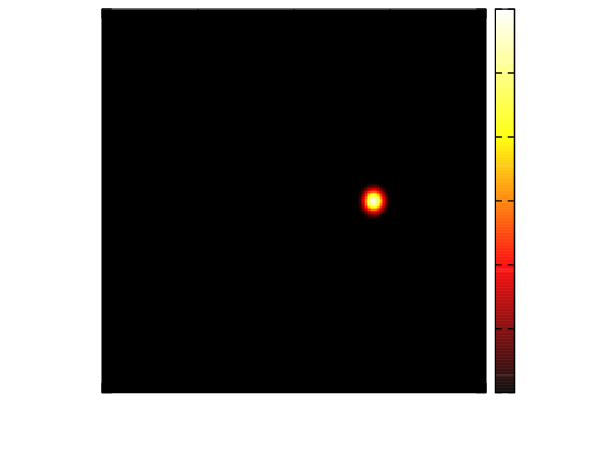}}}
\caption{}
\label{fig:2b}
\end{subfigure}
\begin{subfigure}{0.32\textwidth}
{\resizebox{1\textwidth}{!}{\input{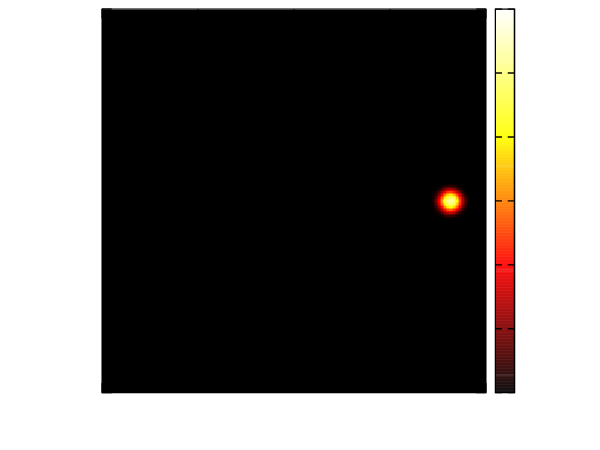}}}
\caption{}
\label{fig:2c}
\end{subfigure}
\begin{subfigure}{0.32\textwidth}
{\resizebox{1\textwidth}{!}{\input{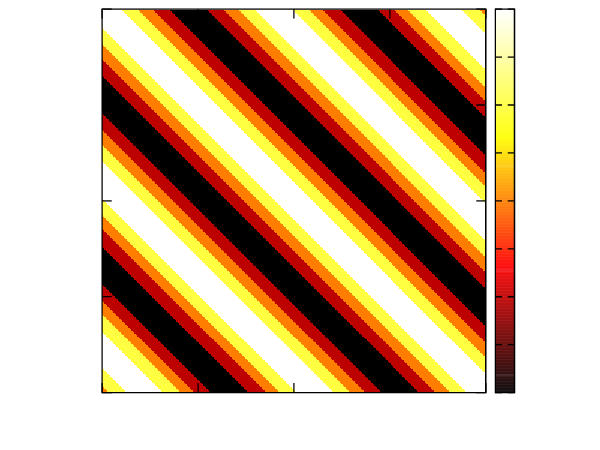}}}
\caption{}
\label{fig:2d}
\end{subfigure}
\begin{subfigure}{0.32\textwidth}
{\resizebox{1\textwidth}{!}{\input{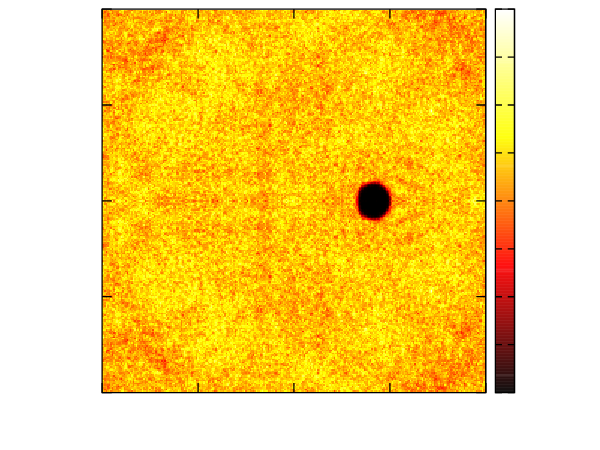}}}
\caption{}
\label{fig:2e}
\end{subfigure}
\begin{subfigure}{0.32\textwidth}
{\resizebox{1\textwidth}{!}{\input{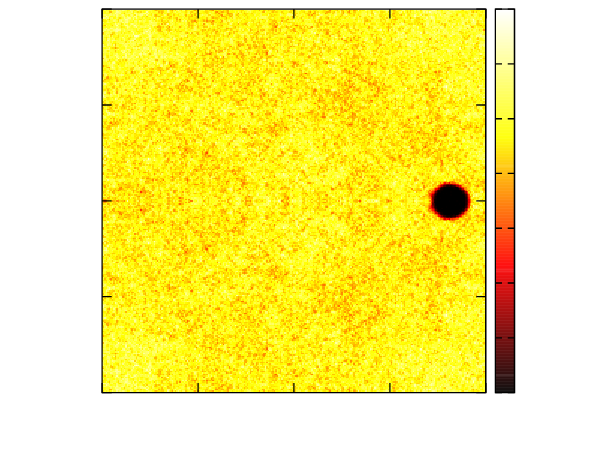}}}
\caption{}
\label{fig:2f}
\end{subfigure}
\caption{Generation of a single DB soliton in the superfluid-impurity system using the initial condition~\eq{IC:1_soliton}. Here, the width of the Gaussian profile is $\sigma=0.5$, and the impurity fraction $\frac{N_2}{N}=0.06$. Upper panel (a)-(c) shows the bright soliton in the impurity component while the lower panel (d)-(f) shows the formation of dark soliton in the superfluid component.}
\label{fig:2}
\end{figure}

One of the dynamical features of these localizations is that, once formed, they propagate with constant velocity, modified only through collisions with each other. Notice that the MI-generated DB solitons appear at random locations and move along random directions. To generate DB solitons in the superfluid-impurity mixture in a controlled manner, we use the following method. We initialize the system by imposing on the continuous wave background of the superfluid component a localized moving Gaussian profile of impurity atoms. For instance, a single 2D moving soliton can be generated by using the wavefunction
\begin{equation}{\label{IC:1_soliton}}
    \psi_2 =\sqrt{\frac{N_2}{\pi\sigma^2}}\ \e^{-\frac{1}{\sigma^2}(x^2+y^2)} \e^{i\phi},
\end{equation}
for impurity atoms along with the homogeneous state, $\psi_1=\sqrt{n_1}$, for the superfluid component. From the hydrodynamic representation~\cite{dalfovo}, the initial velocity of the Gaussian field is determined by the phase $\phi(x,0)$ through the relation $\mathbf{v}=\nabla\phi$. Therefore, to generate a localization with a non-zero velocity, the phase of the initial field must be inhomogeneous in space. The evolution of the Gaussian field, \eq{IC:1_soliton}, with an initial phase
\begin{equation}
    \phi = x,
\end{equation}
is shown figure~\ref{fig:2}. If the dynamics of $\psi_2$ were only governed by \eq{eq:2b}, the initial waveform would eventually spread out due to the dispersive nature of the Schr\"odinger equation (in the presence of a constant potential). However, because of the coupling of impurities to the superfluid component with repulsive intra-component interaction, the Gaussian field evolves into a stable 2d localized structure, resembling a bright soliton (figure(\ref{fig:2})(a)-(c)). 

It is well known that 2D localizations in a single component GPE are unstable due to transverse instability. While in 1D, dispersion is balanced by a combination of  nonlinearity and propagation, in 2D however, dispersion in the transverse direction cannot be completely offset by the cubic nonlinearity alone. In the case of a binary mixture, the stabilization of bright soliton in the impurity component is facilitated by the density minimum formed in the superfluid. The physical mechanism that stabilizes the DB soliton in the superfluid-impurity system can be understood using the mean-field model. To understand the mechanism of localization better we look back at the governing Schr\"{o}dinger equation \eq{eq:2b} for $\psi_2$. While the impurity atoms do not interact with each other (or dilute enough for the interaction potential to be negligible),  the superfluid density 
$|\psi_1|^2$ acts as the potential for the evolution of $\psi_2$. First, the MI of homogeneous state causes superfluid atoms that are already in the unstable mode to destabilize into a density depression. Since the intercomponent interaction is repulsive, superfluid atoms encounter a repulsive barrier created by the gaussian impurity localization and move away from the region, leading to a local dip in its density profile. Any such intensity dip in the superfluid acts as an attractive potential well for the impurity to localize and, when it propagates, is sufficient enough to balance the dispersion in the transverse direction. On the other hand, in \eq{eq:2a}, both $|\psi_1|^2$ and $|\psi_2|^2|$ act as potentials for the evolution  of the superfluid amplitude $\psi_1$, both of which are repulsive. Thus any localization in the impurity component enhances the repulsion between the superfluid atoms, complementing impurity localization further. The transverse dispersion of this density dip is then balanced by the superfluid-impurity coupling, resulting in a stable 2D DB soliton. Figure~(\ref{fig:2}) shows how a dark soliton emerges in the superfluid density from the homogeneous state in the process of forming a DB soliton. 

We note that the localized modes are stable only when the interaction strengths are comparable $\alpha=\frac{g_{12}}{g_{11}}\approx 1$. In contrast, if the superfluid-impurity interaction is stronger, or weaker, than the intracomponent interaction between superfluid atoms, the initial waveform destabilizes and splits into multiple localizations. 

\subsection{Dynamics: Soliton Collisions}
\begin{figure}[tb]
\centering
\begin{subfigure}{0.24\textwidth}
{\resizebox{1\textwidth}{!}{\input{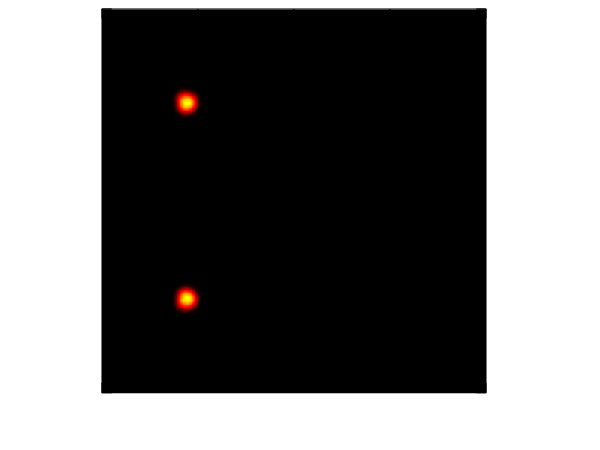}}}
\caption{}
\label{fig:3a}
\end{subfigure}
\begin{subfigure}{0.24\textwidth}
{\resizebox{1\textwidth}{!}{\input{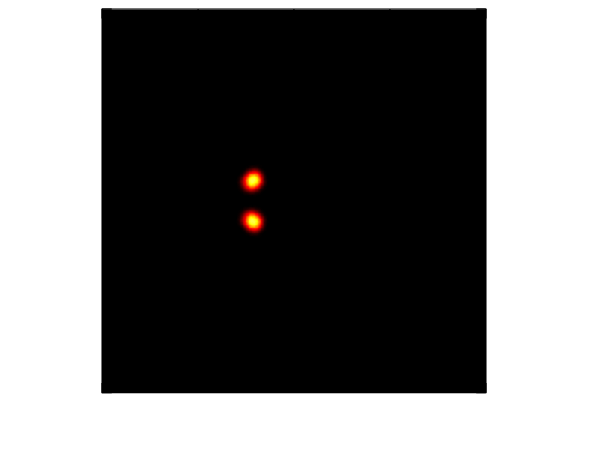}}}
\caption{}
\label{fig:3b}
\end{subfigure}
\begin{subfigure}{0.24\textwidth}
{\resizebox{1\textwidth}{!}{\input{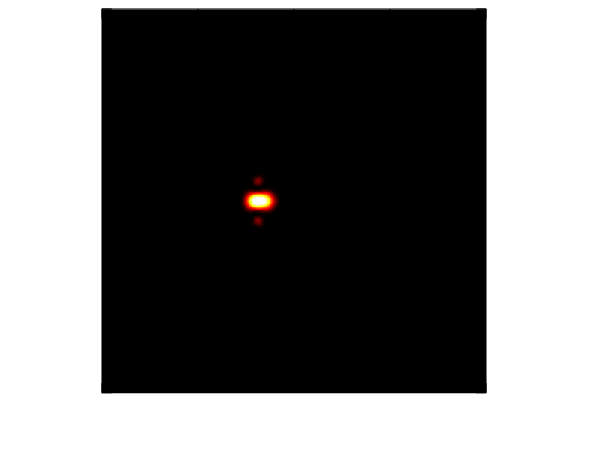}}}
\caption{}
\label{fig:3c}
\end{subfigure}
\begin{subfigure}{0.24\textwidth}
{\resizebox{1\textwidth}{!}{\input{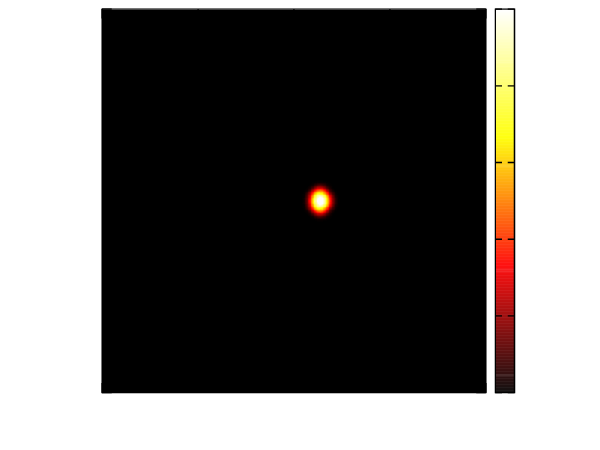}}}
\caption{}
\label{fig:3d}
\end{subfigure}
\\[0.5cm]
\begin{subfigure}{0.24\textwidth}
{\resizebox{1\textwidth}{!}{\input{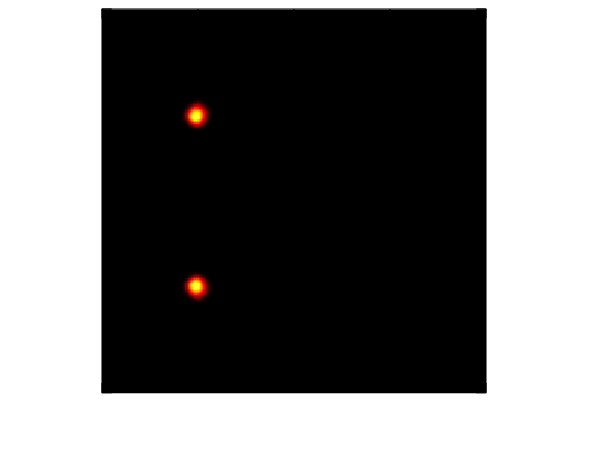}}}
\caption{}
\label{fig:3e}
\end{subfigure}
\begin{subfigure}{0.24\textwidth}
{\resizebox{1\textwidth}{!}{\input{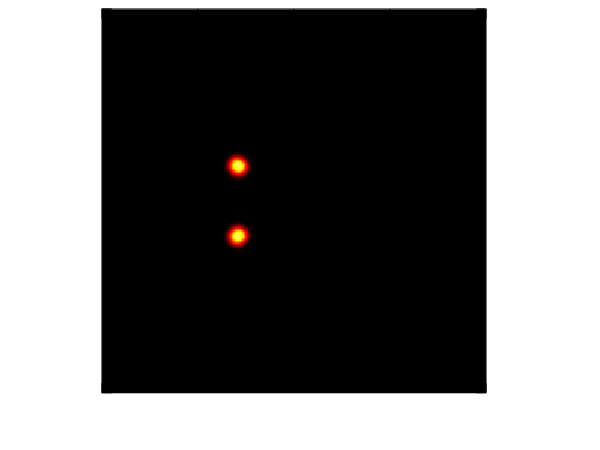}}}
\caption{}
\label{fig:3f}
\end{subfigure}
\begin{subfigure}{0.24\textwidth}
{\resizebox{1\textwidth}{!}{\input{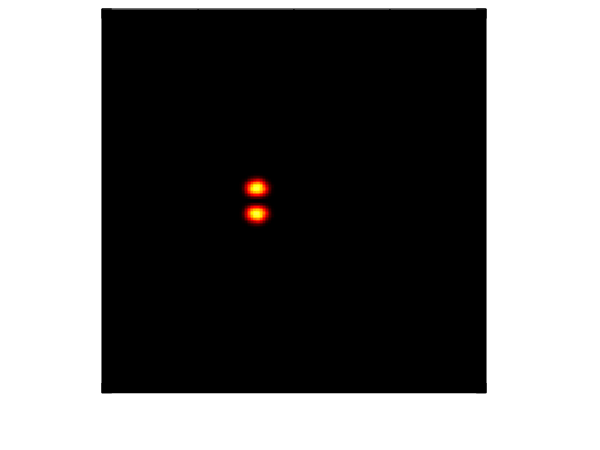}}}
\caption{}
\label{fig:3g}
\end{subfigure}
\begin{subfigure}{0.24\textwidth}
{\resizebox{1\textwidth}{!}{\input{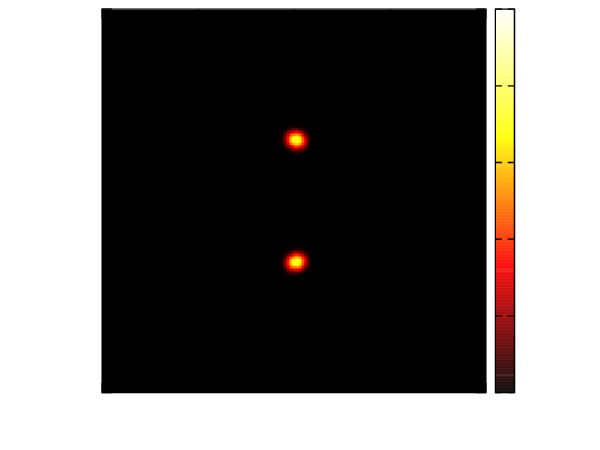}}}
\caption{}
\label{fig:3h}
\end{subfigure}
\caption{Soliton collisions in the superfluid-impurity system for repulsive interactions: Upper panel(a)-(d) Merging of solitons when the phase difference between individual Gaussian wavepackets is zero. Lower panel (e)-(h) Soliton repulsion when the phase difference between initial wavepackets is $\pi$. In both cases, the impurity component is a linear superposition of two Gaussian wavefunctions  symmetrically placed about the x-axis with identical amplitude profiles as given by~\eq{eq:collision}. For a detailed animation of merging and repulsion, see supplementary material. }
\label{fig3}
\end{figure}

\begin{figure}[tb]
\centering
\subfloat[]%
{\resizebox{0.48\textwidth}{!}{\input{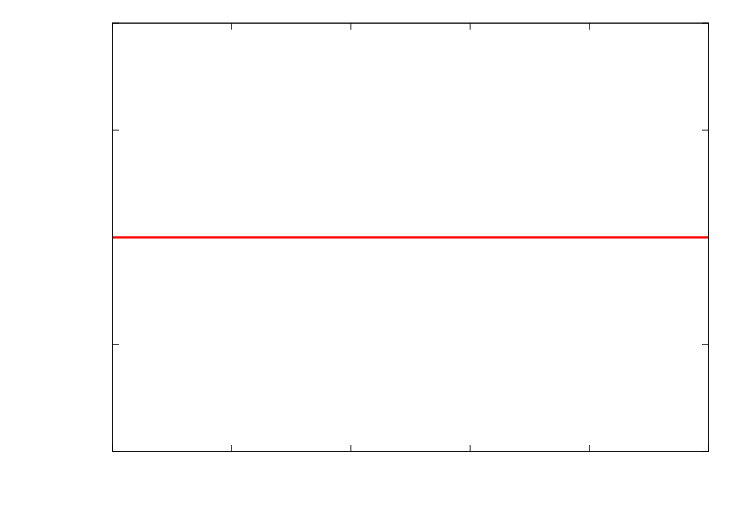}}}
\label{fig4a}
\hfill
\subfloat[]%
{\resizebox{0.48\textwidth}{!}{\input{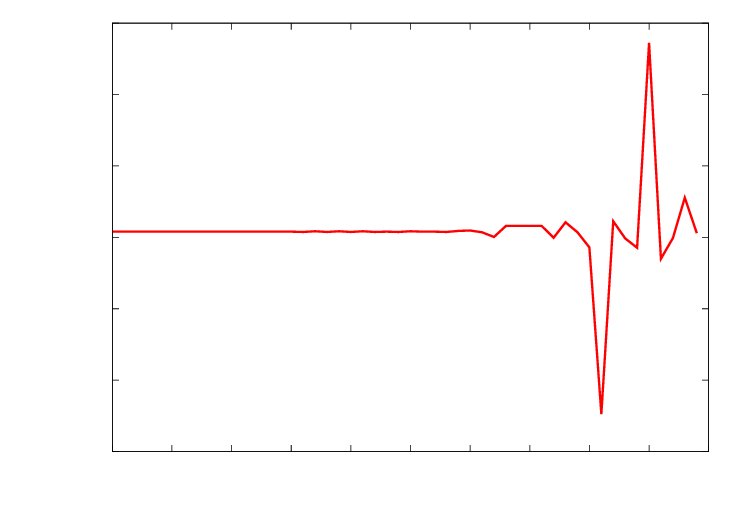}}}
\label{fig4b}
\caption{Relative phase between the bright components of DB solitons as a function of time when the initial phase difference between the pulses is (a) zero (merging of solitons) and (b) $\pi$ (repelling solitons).}
\label{fig4}
\end{figure}

The method of generation of a single soliton discussed in the previous section can be extended to study the dynamics of multiple localizations. Here we investigate the dynamics of DB soliton collisions in the binary mixture when the inter- and intra-component interactions are repulsive. To generate two DB solitons, we consider an initial state where the impurity component consists of two non-overlapping Gaussian wavepackets on a homogeneous background of superfluid density:
\begin{equation}{\label{eq:collision}}
    \psi_2 = \psi^{(1)}_2+\psi^{(2)}_2.
\end{equation}
The wavepackets
\begin{equation}
    \psi^{(j)}_2= A_j\e^{-\frac{1}{\sigma^2}((x-x_j)^2+(y-y_j)^2)} \e^{i\phi_j},
\end{equation}
are assumed to be identical Gaussian profiles located at $(x_j,y_j)$, and whose widths are smaller compared to the distance between them to avoid overlap. Each packet is initialized with a phase as in~\eq{IC:1_soliton}, which also determines its propagation velocity. Figure~\ref{fig3} shows two Gaussian wavepackets initialized symmetrically about the x-axis and colliding with each other. Although the underlying interactions between atoms are repulsive, DB solitons can exhibit both coalescence and repulsion during collisions. This behavior is strongly dependent on the relative phase between the bright components. In particular, we observe that when the initial phase difference between the localizations is zero, they fuse together upon collision to form a single matter-wave soliton. In figure~\ref{fig3}(a)-(d), we show the interaction between DB solitons generated using two identical Gaussian wavefunctions placed symmetrically about the x-axis at $(-8,-8)$ and $(-8,8)$. The individual phases of the initial waveforms are $\phi_1=\frac{x+y}{2}$ and $\phi_2 = \frac{x-y}{2}$, respectively, such that the phase difference $\Delta \phi = \phi_2-\phi_1$ between the centers is zero. Since the widths of the solitons are small, we shall approximate the phase difference between them using the center of localizations. Interestingly, we observe that the phase difference between the bright components remains constant until they coalesce (figure~(\ref{fig4}a)). 

\begin{figure}[htb]
\centering
\begin{subfigure}{0.48\textwidth}
    {\resizebox{1\textwidth}{!}{\input{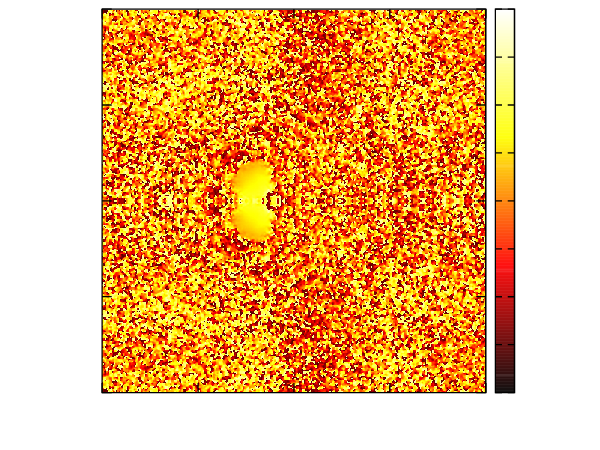}}}
    \caption{}
\end{subfigure}
\begin{subfigure}{0.48\textwidth}
    {\resizebox{1\textwidth}{!}{\input{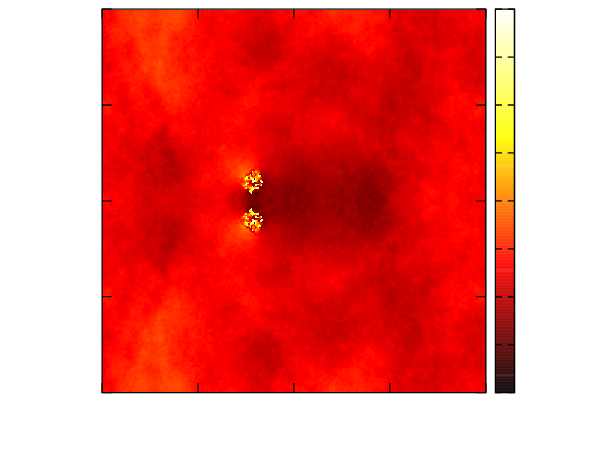}}}
    \caption{}
\end{subfigure}
\\[1.2cm]
\begin{subfigure}{0.48\textwidth}
    {\resizebox{1\textwidth}{!}{\input{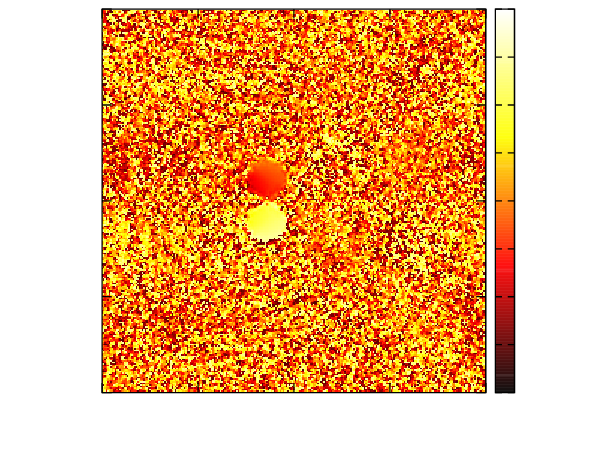}}}
    \caption{}
\end{subfigure}
\begin{subfigure}{0.48\textwidth}
    {\resizebox{1\textwidth}{!}{\input{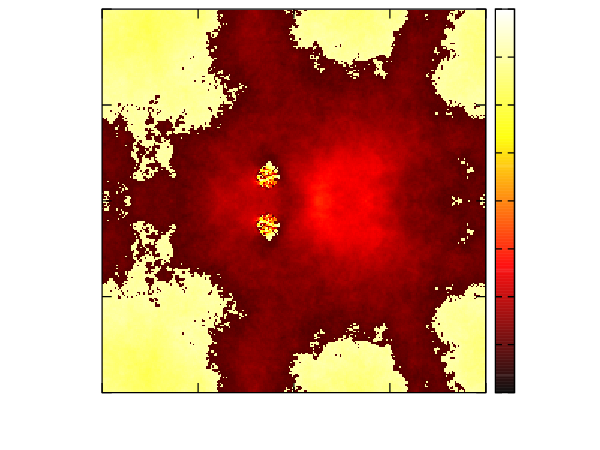}}}
    \caption{}
\end{subfigure}
\caption{Phase plots of impurity ((a) and (c)) and  superfluid ((b) and (d)) components just  before (a),(b) the two DB solitons coalesce and (c), (d)  they repel each other.}
\label{fig:phase}
\end{figure}

As shown in figure~\ref{fig3}(e)-(h), when the phase difference between the localizations is $\pi$, the DB solitons repel each other. In this case, the phases of individual bright components vary, yet the relative phase is approximately constant. In a non-integrable system, such a phase coherence is not long-lived, and successive collisions between the solitons change the relative phase. For intermediate values of $\Delta \phi$, the interacting solitons attract if $\Delta \phi$ lies in the range $0<\Delta\phi<\frac{\pi}{2}$ and repel if $\frac{\pi}{2}<\Delta\phi<\pi$. In contrast to an in-phase or out-of-phase collision, these interactions do not preserve the relative phase. The complete phase profile of the components before collision is shown in figure~\ref{fig:phase}. The phase profile of the superfluid components indicates that the density depressions are not vortices.

The apparent attraction or repulsion between solitons in the mean-field model is because of the dependence of the density of impurity atoms on the relative phase. To illustrate this, we consider the wavefunction of the impurity component for two interacting solitons
\begin{equation}
    \psi_2 = \rho_1\e^{i\phi_1}+\rho_2 \e^{i\phi_2},
\end{equation}
where $\rho_j$ and $\phi_j$ are the amplitudes and phases of the individual solitons. For two well-separated solitons, the density of impurity atoms, which determines the stability of DB solitons, is simply the sum of individual contributions:

$$|\psi_2|^2\approx\rho_1^2+\rho_2^2.$$
 
However, in the region of overlap, the superposition of the two solitons modifies the inter-component interaction in~\eq{eq:2a} due to the additional phase dependent term 
\begin{equation}
    |\psi_2|^2 = \rho_1^2+\rho_2^2+2\rho_1\rho_2\cos(\Delta\phi).
\end{equation}
For an in-phase collision, overlap between the solitons $|\psi_2|^2= (\rho_1+\rho_2)^2$ is enhanced, resulting in a stronger repulsive superfluid-impurity coupling compared to the well-separated case. This increased repulsion causes the superfluid atoms to move away from the region of overlap, effectively leading to coalescence of solitons. Conversely, for an out-of-phase collision between two solitons, the density $|\psi_2|^2=(\rho_1-\rho_2)^2$ is suppressed in the region of overlap, which reduces the superfluid-impurity repulsion. This weaker interspecies repulsion allows more superfluid atoms to occupy the region of overlap, resulting in an effective repulsion between the solitons. In other words, the phase difference between the bright solitons induces an effective attractive/repulsive interaction between the density depressions (dark solitons) of the superfluid component. 

\section{Discussion and Conclusions}
\label{sec:5}
In a single component NLSE, a dark soliton excitation in two dimensions destabilizes into a vortex state due to transverse instabilities, naturally favoring a topological excitation over a conventional soliton. Thus, two dimensional solitons are difficult to sustain in single component NLSEs. In contrast, the dynamical model presented here demonstrates that two dimensional coupled BECs can support stable localizations in the form of DB solitons under a physically realizable limiting case with repulsive intra- and inter-component interactions. In this setting, the MI of the continuous wave solutions induces phase separation in the binary BECs resulting in the formation of localized DB solitons. Unlike vortex solutions, the DB solitons discussed in this work are non-topological localizations stabilized by inter-component interactions. This distinction highlights the potential of coupled BECs as a physical platform for investigating interaction between topological vortices and DB solitons.

In summary, we have investigated the dynamics of symbiotic DB solitons in the superfluid-impurity system with repulsive contact interactions. The absence of self-interactions in the impurity component on account of its diluteness favors clustering of impurity atoms into stable localizations which are effectively trapped by the density dips in the superfluid component. The density depletions in the superfluid component are not vortices, as confirmed by plotting the phase of the corresponding wave function. Moreover, the dynamics of DB solitons shows merging and repulsion determined by their relative phase.

\end{document}